\documentstyle[prl,aps,psfig,fleqn,epsfig,axodraw,alltt,xspace,twocolumn]{revtex}

\def\GeV{\text{GeV}} 
\def\centim{\text{cm}} 
\def\GeVc2{\text{GeV/$c^2$}}
\def\mrad{\text{mrad}} 
\def\meter{\text{m}} 
\def\kmeter{\text{km}}
\def\cmeter{\text{cm}}
\def\kGauss{\text{kG}}
\def\gccm{\text{g/cm}^3}

\def\tanb{$\tan \! \beta$}
\def\chizero{$\tilde{\chi}^0_1$\xspace}
\def\mchizero{m_{\tilde{\chi}^0_1}}

\def\Rpar{$R$-parity\xspace}
\def\Rpv{\mbox{$\not \!\! R_p$}\xspace}

\def\lijk{$\lambda_{ijk}$\xspace}
\def\lpijk{$\lambda^\prime_{ijk}$\xspace}
\def\lppijk{$\lambda^{\prime\prime}_{ijk}$\xspace}

\newlength{\defaultparindent}
\begin{document}

\draft

\title{Searching for the Lightest Neutralino at Fixed Target
Experiments}

\author{
L. Borissov, J. M. Conrad, M. Shaevitz
}

\address{
Columbia University, New York, NY, 10027
}

\date{\today}

\maketitle

\begin{abstract}
\par Most ongoing supersymmetry searches have concentrated on
the high-energy frontier. High-intensity fixed target beamlines, however, offer
an opportunity to search for supersymmetric particles with long lifetimes and low
cross-sections in regions complementary to the ones accessible to collider
experiments. In this paper, we consider \Rpar violating supersymmetry searches
for the lightest neutralino and use the NuTeV experiment as an example for the
experimental sensitivity which can be achieved.

\end{abstract}


\section{Motivation}
\par

A review of sypersymmetric models can be found in Ref. \cite{Martin:1997ns}. We
consider models where the lightest neutralino~(\chizero) is the Lightest
Supersymmetric Particle (LSP). If the LSP is allowed to decay, \Rpar violation
(\Rpv) is required via the superpotential:

\begin{equation}
W_{\not R_{p}}=\lambda_{ijk}L_{i}L_{j}\bar{E}_{k}
+\lambda ^{_{\prime }}_{ijk}L_{i}Q_{j}\bar{D}_{k}
+\lambda ^{_{\prime \prime }}_{ijk}\bar{U}_{i}\bar{D}_{j}\bar{D}_{k}
\label{Wrpveq}
\end{equation}
where $i$, $j$ and $k$ are generation indices, $L$ and $Q$ are the lepton and
quark $SU(2)$ superfield doublets, $E$, $U$, and $D$ are the lepton and quark singlets
and \lijk, \lpijk, and \lppijk are Yukawa-type couplings between the
fields. This model is specified by the squark and slepton masses; mass terms for
the gauginos at the electroweak scale ($M_1$, $M_2$ and $M_3$), the ratio of
vacuum expectation values of the two neutral Higgses (\tanb); a mass term mixing
the two Higgs doublets ($\mu$) and the values of the $\lambda$-couplings.

\par 
In the Minimal Supersymmetric Standard Model (MSSM), unification is imposed at the
GUT-scale, which leads to the relation:

\begin{equation}
$
\(M_1=\frac{5}{3} \tan^2\theta_W M_2 \approx 0.5~M_2\)
$
\label{guteq}
\end{equation}
where $\theta_W$ is the weak mixing angle at the
electroweak scale. Assuming Eq.~(\ref{guteq}), current LEP data give
\mbox{$\mchizero > 32.3~\GeV$} \cite{Abreu:1999qz,ALEPH}. In the \Rpv scenario,
collider experiments require the \chizero to decay inside the detector. This
leads to very low sensitivity for a long lifetime low mass \chizero
\cite{Abreu:1999qz}. Moreover, if the GUT-scale unification requirement
(\ref{guteq}) is not assumed, such neutralinos can be produced in observable
quantities in lifetime regions inaccessible to collider experiments. Coverage in
these regions can be achieved by a detector far from the collision vertex,
provided there is enough luminosity and cross-section for neutralino production.

\par In the following analysis, we present an example of the experimental
sensitivity for \Rpv SUSY with lepton number violation ($\lambda_{ijk}> 0$) of a
high-intensity fixed target experiment. We are interested in very low mass
\chizero, so we do not require the relation given in Eq.~(\ref{guteq}), leaving
$M_1$ and $M_2$ to be free parameters. However, bounds from the invisible decay
of the $Z^0$ necessitate that $M_1$ is very small compared to $M_2$, so that 
the \chizero is mostly
bino and only its Higgsino admixture couples to the $Z^0$
\cite{Caso:1998tx,Haber:1985rc}. Thus we work in the framework of
phenomenologically motivated \Rpv unconstrained MSSM (uMSSM). However our model
is very close to the MSSM except for GUT-scale unification and \Rpar
conservation.

\section{An example: NuTeV} 

\par The NuTeV experiment (E815) \cite{Nutev1,Nutev2,Nutev3} took data during the
1996-1997 fixed target run at Fermilab. NuTeV was designed to perform precision
measurements of Standard Model parameters through deep inelastic neutrino-nucleon
scattering. These measurements provide competitive and in many cases unique tests
of the present-day understanding of the Standard Model, complementary to direct
measurements from collider experiments.

\par During the run, $800~\GeV$ protons from the Tevatron are incident upon a
$30.5~\centim$ BeO target at a $7.8~\mrad$ angle. A total of $2.86 \! \times \!
10^{18}$ protons on target were recorded during the live time of the detector.
The protons interact and produce a secondary beam of pions, kaons and other
hadrons. A system of quadrupole magnets -- the Sign-Selected Quadrupole Train
(SSQT) picks out secondaries of the correct sign and dumps wrong sign particles.
After focusing, the pions and kaons enter a $541~\meter$ decay pipe, where they
decay in flight with dominant modes \mbox{$\pi \rightarrow \nu_\mu \mu$},
\mbox{$K \rightarrow \nu_\mu \mu$} and \mbox{$K \rightarrow \pi^0 \nu_\mu \mu$}.
Approximately $3\%$ of the mesons decay in the pipe, the rest are dumped in
$6~\meter$ of aluminum and steel. The muons range out in $241~\meter$ of steel
shielding and $582~\meter$ of earth berm. The resulting beam is a nearly pure
neutrino beam with less than $0.2\%$ contamination; it exhibits a dichromatic
spectrum corresponding to the superposition of $\pi$ and $K$ decay distributions.

\par
\begin{figure}[ht]
\centerline{\epsfig{figure=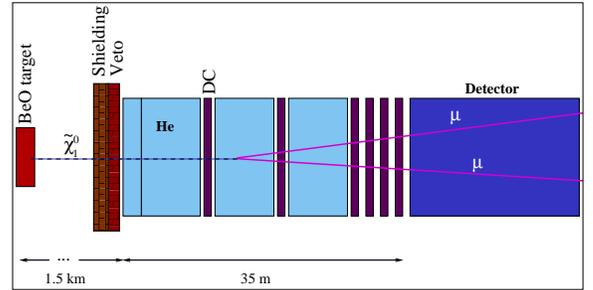,width=3.in}}
\caption{The NuTeV decay channel. A possible 
$\tilde{\chi}^0_1\rightarrow \nu_e \mu^+ \mu^-$ decay 
is shown as seen by the detector.}  
\label{hebagsfig}
\end{figure}

 The NuTeV detector is situated $1.5~\kmeter$ downstream from the target. The
calorimeter consists of 168 steel plates
($3~\meter\times3~\meter\times5.1~\cmeter$), 84 liquid scintillation counters
(every $10.2~\cmeter$ of steel) and 42 drift chambers (every $20.4~\cmeter$ of
steel) followed by a $15~\kGauss$ toroidal spectrometer. Upstream of the
calorimeter is a $35~\meter$ decay region consisting of three large helium bags
and six drift chambers (Fig.~\ref{hebagsfig}). The decay channel is shielded by
an upstream veto wall.

\subsection{Production and decay}

\par
At the NuTeV target, \chizero's can be pair-produced in the $s$-channel via
a $Z$, or in the $t$-channel through an exchange of a squark
(Fig.~\ref{prodfig}). If the squark mass is sufficiently small, 
production can be enhanced. For our estimates, however,
we have chosen to work with sfermion masses of the order of $800~\GeV$,
conservatively above present experimental bounds \cite{Caso:1998tx}.
Thus the only relevant production parameters
are $M_1, M_2, \mu$, and \tanb. Production is insensitive to $M_3$,
which is responsible for the gluino mass.

\par
NuTeV uses a high intensity proton beam, but its center-of-mass
energy ($\sqrt s \approx 39~\GeV$) is low compared to collider
experiments. This, however, simplifies the search, since the only
relevant mode at this energy is neutralino pair production.

\par
In \Rpv MSSM, the \chizero decays leptonically according to the diagrams shown in
Fig.~\ref{chidecayfig}. It is often assumed that only one of the $LLE$ operators
in Eq.~(\ref{Wrpveq}) has a sizeable \lijk coupling, in order to explain the
smallness of $L$-violating terms in the renormalizable Lagrangian required by
present limits \cite{Barger:1989rk}. This is also the case of greatest
experimental interest. For example, if $\lambda_{122}$ is dominant, the \chizero
decays into \mbox{$\nu_e\mu^+\mu^-$} and \mbox{$\nu_\mu e^\pm \mu^\mp$} with
branching ratios of approximately $50\%$ in each of the two channels
\cite{Abreu:1999qz}. The final state of two leptons and missing energy can be
easily detected by an experiment such as NuTeV. The other extreme case,
$\lambda_{133}$ dominant, leads to decays mainly into taus and electrons. The
efficiencies for the other \lijk couplings lie between these two cases.

\par 
The partial width for the \chizero decaying into $\nu ll^\prime$ is of
the typical form for a fermion three-body decay
\cite{Barger:1989rk}:

\begin{equation}
$
\( \Gamma _{\tilde{\chi }^0_1}=
K^{2}\frac{G^{2}_{F}~m^5_{{\tilde{\chi }^0_1}}}{192~ \pi ^{3}} \)
$
\label{decaywidtheq}
\end{equation}
where $K$ is an effective four-fermion coupling in \chizero decays
(Fig.~\ref{chidecayfig}) proportional to the \chizero$f\tilde{f}$
coupling and the \Rpv coupling \lijk. For a large region of SUSY
parameter space

\begin{equation}
$
\( K \sim 0.1 \biggl({\frac{100~\GeV}{m_{\tilde{f}}}}\biggr)^2 \lambda_{ijk} \)
$
\label{Keq}
\end{equation}
where $m_{\tilde{f}}$ is the mass of the virtual selectron or
sneutrino exchanged \cite{Haber:1985rc}. 

\par
Upper bounds on \lijk and can be found in Refs.
\cite{Barger:1989rk,Allanach:1999ic}. The ones pertinent to our case
of study are:

\begin{equation}
\lambda_{122} < 0.049 \times \frac{m_{\tilde{e}_R}}{100~\GeV}
\label{l122eq}
\end{equation}

\begin{equation}
\lambda_{133} < 0.006 \times \sqrt{\frac{m_{\tilde{\tau}}}{100~\GeV}}
\label{l133eq}
\end{equation}
which come from current universality requirements and limits on
$\nu_e$ mass \cite{Allanach:1999ic}.

\par
Eq. (\ref{Keq}) can be rewritten in terms of the average decay length in the
lab frame:

\begin{equation}
l (\text{cm}) =0.3(\beta\gamma){\biggl( \frac{m_{\tilde{f}}}{100~\GeVc2} \biggr)}^4
{\biggl( \frac{1~\GeVc2}{m_{\tilde{\chi}^0_1}} \biggr)}^5 \frac{1}{\lambda^2}
\label{Leq}
\end{equation}

\subsection{Predictions for the NuTeV experiment}

\par
The expected number of \chizero decays detectable by NuTeV
is the product of the number \chizero's produced at the target passing
through the detector and the probability that any of them actually
decays in the He decay region (Fig.~\ref{hebagsfig}). 

\begin{equation}
N_{evt}={\biggl( N_{p}~d~N_{A}~ 
\rho_{BeO}\frac{d\sigma } {d\Omega }d\Omega \biggr)}
\times \frac{(1-e^{-\Delta z/l})} {e^{z/l}}
\label{nodeq}
\end{equation}
where $N_p=2.8 \times 10^{18}$ is the the number of protons on target,
$d$ is the length of the BeO target, $\rho_{BeO}=2.7~\gccm$ is the
target mass density, $N_A$ is Avogadro's number; $d\sigma/d\Omega$ is the
differential production cross-section in the center-of-mass reference
frame of the pair-production, \mbox{$d\Omega \approx 6.7 \times
10^{-3}~\text{srad}$} is the solid angle subtended by the 
detector in that reference frame; $z$ is the distance between the
target and the decay region, $\Delta z$ is the length of the decay
region and $l=\gamma c \tau$ is the \chizero decay length in
the lab frame. For the NuTeV specifications, equation (\ref{nodeq}) gives: 

\begin{equation}
N_{evt}\approx
\biggl(2.4 \times 10^{14}\text{mb}^{-1}\frac{d\sigma}{d\Omega } \biggr)
\times 
\frac {1-e^{ \frac{-3.5 \times 10^3\cmeter}{l}}}
{e^{ \frac{1.5 \times 10^5\cmeter}{l}}}
\label{calceq}
\end{equation}

\par

\begin{figure}[ht]
\centerline{\epsfig{figure=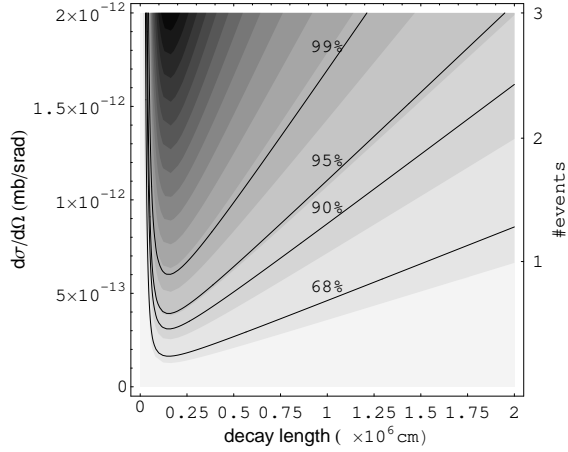,width=3.in}}
\caption{Contour plots of \chizero pair-production differential
cross-section seen by the NuTeV detector versus \chizero decay
length in the lab frame. The shaded contours represent the number of 
\chizero decays occuring inside the detector assuming perfect detector 
efficiency. The four confidence level contours represent 
exclusion regions (above the contours) following a Feldman and Cousins approach for
0 signal events and 0 background [20].}   
\label{nutevclfig}
\end{figure}

From Eq.~(\ref{calceq}), we make a contour plot of the expected
number of observable \chizero decays versus $d\sigma/d\Omega$ and $l$ 
as shown in Fig.~\ref{nutevclfig}. Since the expected signal is
small, we use a Feldman and Cousins approach for the analysis
\cite{Feldman:1998qc}. The confidence level contours
obtained are shown on Fig.~\ref{nutevclfig}. 
We continue to work with the decay length in the lab frame, since it 
conveniently folds the uncertainty in \lijk, the sfermion mass and the
the \chizero mass, which are not very rigidly constrained by present
SUSY searches. Moreover, the range of $l$ can be estimated by direct
experimental
observation. If such an experiment sees a non-zero signal, phenomenological
information about the \chizero mass can be incorporated to constrain the 
remaining SUSY parameters.

As an example of the experimental sensitivity, we consider the hypothetical case
of 0 events, i.e. very low signal. We assume 0 background, which is consistent
with background estimates for the NuTeV experiment \cite{Formaggio:1999ne}.

\par
\begin{figure}[ht]
\centerline{\epsfig{figure=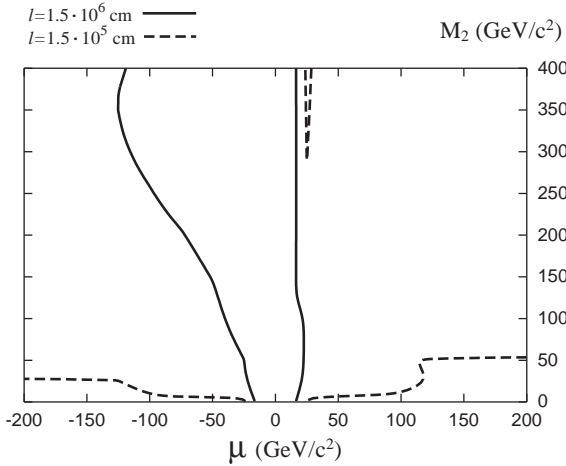,width=3.in}}
\caption{Sensitivity for the NuTeV experiment  for 
small \tanb~$=1.5$ at $M_1 = 1~GeV/c^2$ for two representative
decay lengths. Regions in uMSSM space inside
 the contour lines are
excluded at 90\% CL  using a Feldman and Cousins approach (see Fig \ref{nutevclfig}).}

\label{smalltb}
\end{figure}
\begin{figure}[hb]
\centerline{\epsfig{figure=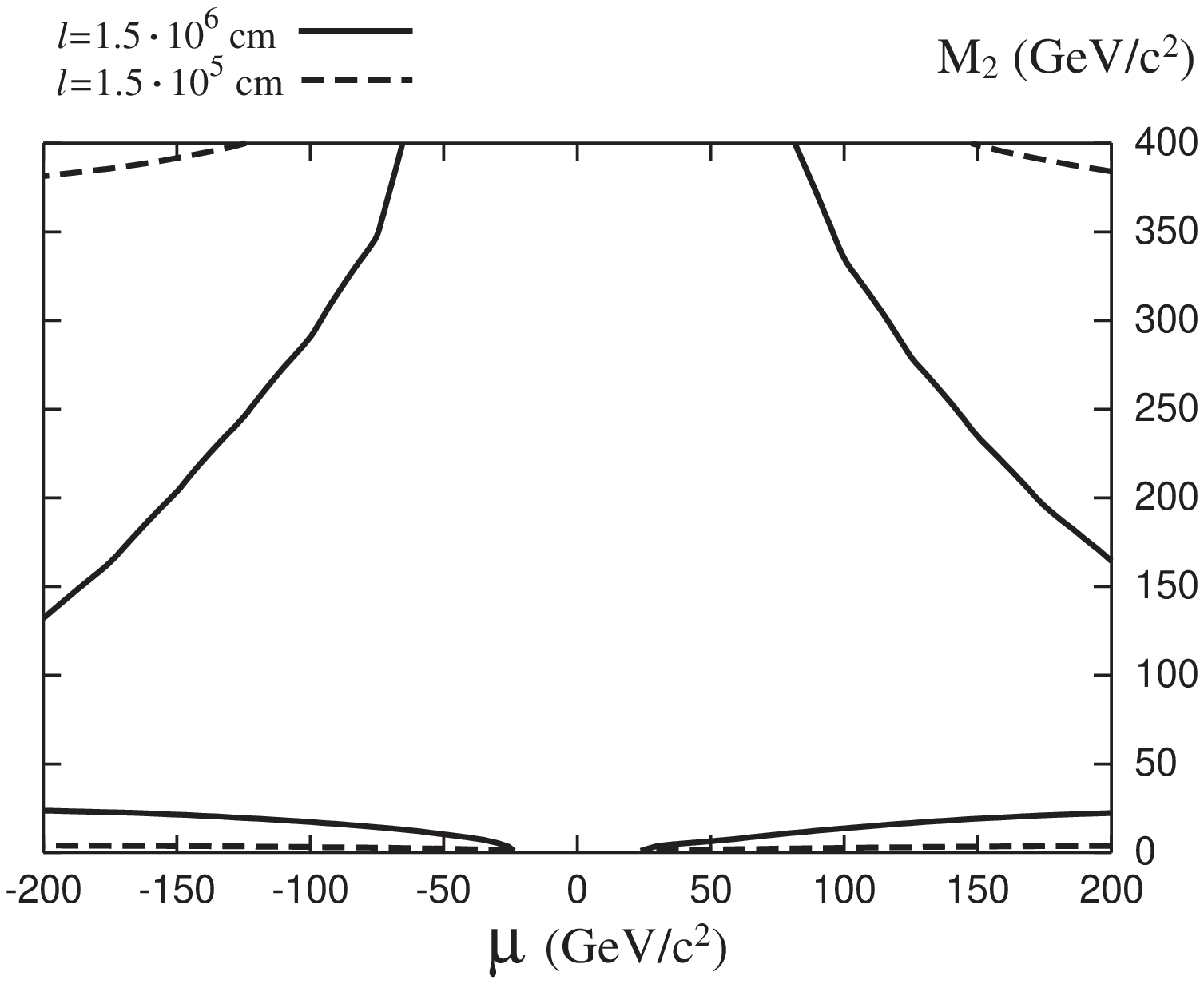,width=3.in}}
\caption{Sensitivity for the NuTeV experiment for 
large \tanb~$=30$ at $M_1 = 1~GeV/c^2$. Regions in uMSSM space inside
 the contour lines are
excluded at 90\% CL  using a Feldman and Cousins approach (see Fig \ref{nutevclfig}).}

\label{largetb}
\end{figure}

Using a Monte Carlo event generator \cite{Mrenna:1997hu,Sjostrand:1995iq}, 
we perform a scan of uMSSM parameter space for 
$M_1=1,10,100~\GeVc2$, $M_2=0,...,400~\GeVc2$,
$\mu = -200,...,200~\GeVc2$ and $\text{tan} \beta = 1.5,...,40$. We set 
$m_{\tilde{f}} \approx 800~\GeVc2$ and \mbox{$M_3 = 1~\text{TeV}$}. We found no strong 
dependence on $M_1$ 
and $m_{\tilde{f}}$, as long as \mbox{$M_1 << M_2$} so that \chizero is mostly bino.
We consider \mbox{$M_1=1~\GeVc2$} as indicative for the case accessible to NuTeV and
similar fixed target experiments and present results
for two representative decay lengths: \mbox{$l=1.5\times10^5 \text{cm}$}, where 
sensitivity is optimal, 
and \mbox{$l=1.5\times10^6 \text{cm}$}, far past the detector, where sensitivity is
 significantly reduced.
 The results for small and 
large \tanb~ are shown 
on Fig. \ref{smalltb} and Fig. \ref{largetb} respectively.  The exclusion regions in these
plots are similar to the ones obtained by the LEP experiments \cite{Abreu:1999qz} with 
the difference that the latter address the case when the
\chizero decays inside the LEP detectors while we consider the case of large \chizero
decay length, well beyond any detector placed at the production vertex.

\section{Conclusions}
\par
The work presented here is an attempt to motivate \Rpv searches at
fixed target experiments, since collider experiments
run into sensitivity problems at low energy
\cite{Caso:1998tx,Abreu:1999qz}. We argue that NuTeV and 
similar fixed target experiments, such as KTeV, may be in a unique
position to complement collider searches. Recently, the work presented 
in this paper has been applied by the NuTeV collaboration to set a limit 
on neutralino production \cite{Adams:2000rd}.

\section{Acknowledgements}
\par
We would like to thank to C. Quigg, J. Lykken, M. Carena, V. Barger, L. DiLella, 
P. Nienaber and the
 NuTeV collaboration. This research was supported by the U.S. Department of 
Energy and the National Science Foundation.

\begin{figure}[hb]
\centerline{
\unitlength=1.0 pt
\SetScale{1.0}
\SetWidth{0.7}      
\scriptsize    
{} \qquad\allowbreak
\begin{picture}(95,79)(0,0)
\Text(15.0,70.0)[r]{$q$}
\ArrowLine(16.0,70.0)(37.0,60.0) 
\Text(15.0,50.0)[r]{$\bar{q}$}
\ArrowLine(37.0,60.0)(16.0,50.0) 
\Text(47.0,61.0)[b]{$Z$}
\DashLine(37.0,60.0)(58.0,60.0){3.0} 
\Text(80.0,70.0)[l]{$\tilde{\chi}^0_1$}
\Line(58.0,60.0)(79.0,70.0) 
\Text(80.0,50.0)[l]{$\tilde{\chi}^0_1$}
\Line(58.0,60.0)(79.0,50.0) 
\end{picture} \ 
{} \qquad\allowbreak
\begin{picture}(95,79)(0,0)
\Text(15.0,70.0)[r]{$q$}
\ArrowLine(16.0,70.0)(58.0,70.0) 
\Text(80.0,70.0)[l]{$\tilde{\chi}^0_1$}
\Line(58.0,70.0)(79.0,70.0) 
\Text(54.0,60.0)[r]{$\tilde{q}$}
\DashArrowLine(58.0,70.0)(58.0,50.0){1.0} 
\Text(15.0,50.0)[r]{$\bar{q}$}
\ArrowLine(58.0,50.0)(16.0,50.0) 
\Text(80.0,50.0)[l]{$\tilde{\chi}^0_1$}
\Line(58.0,50.0)(79.0,50.0) 
\end{picture} \ 
}
\caption{Neutralino pair production mechanisms.}
\label{prodfig}
\end{figure}
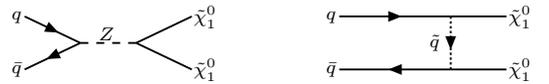

\begin{figure}[ht]
\centerline{
\unitlength=1.0 pt
\SetScale{1.0}
\SetWidth{0.7}      
\scriptsize    
{} \qquad\allowbreak
\begin{picture}(95,79)(0,0)
\Text(15.0,60.0)[r]{$\tilde{\chi}^0_1$}
\Line(16.0,60.0)(58.0,60.0) 
\Text(80.0,70.0)[l]{$\nu$}
\Line(79.0,70.0)(58.0,60.0) 
\Text(54.0,50.0)[r]{$\tilde{\nu}$}
\DashLine(58.0,60.0)(58.0,40.0){1.0} 
\Text(80.0,50.0)[l]{$l^+$}
\Line(58.0,40.0)(79.0,50.0) 
\Text(80.0,30.0)[l]{$l^{\prime-}$}
\Line(79.0,30.0)(58.0,40.0) 
\end{picture} \ 
{} \qquad\allowbreak
\begin{picture}(95,79)(0,0)
\Text(15.0,60.0)[r]{$\tilde{\chi}^0_1$}
\Line(16.0,60.0)(58.0,60.0) 
\Text(80.0,70.0)[l]{$l^\pm$}
\Line(58.0,60.0)(79.0,70.0) 
\Text(54.0,50.0)[r]{$\tilde{l}^\mp$}
\DashLine(58.0,40.0)(58.0,60.0){1.0} 
\Text(80.0,50.0)[l]{$\nu$}
\Line(79.0,50.0)(58.0,40.0) 
\Text(80.0,30.0)[l]{$l^{\prime\mp}$}
\Line(79.0,30.0)(58.0,40.0) 
\end{picture} \ 
{} \qquad\allowbreak
}
\caption{\Rpv \chizero decay mechanisms.}  
\label{chidecayfig}
\end{figure}

\pagebreak

\par

\end{document}